\documentclass[pra,showpacs,amsmath,amssymb,onecolumn]{revtex4}
\usepackage{amssymb}
\usepackage{mathrsfs}
\usepackage{amsfonts}

\usepackage{amsthm}
\usepackage{dcolumn}
\usepackage{bm}
\usepackage{graphicx}

\newcommand\ket[1]{\ensuremath{|#1\rangle}}
\newcommand\bra[1]{\ensuremath{\langle#1|}}

\newcommand\oprod[2]{\ensuremath{|#1\rangle\langle#2|}}

\begin{document}

\title{Unified Approach to Universal Cloning and Phase-Covariant Cloning}

\author{Jia-Zhong Hu}
  \email{hjz07@mails.tsinghua.edu.cn}
  \affiliation{Department of Physics, Tsinghua
University, Beijing 100084, China }
\author{Zong-Wen Yu}
  \email{yzw04@mails.tsinghua.edu.cn}
  \affiliation{Department of Mathematical Sciences, Tsinghua
University, Beijing 100084, China }
\author{Xiang-Bin Wang}
  \email{xbwang@mail.tsinghua.edu.cn}
  \affiliation{Department of Physics, Tsinghua
University, Beijing 100084, China}

\begin{abstract}
We analyze the problem of approximate quantum cloning when the
quantum state is between two latitudes on the Bloch's sphere. We
present an analytical formula for the optimized 1-to-2 cloning. The
formula unifies  the universal quantum cloning (UQCM)
 and the phase covariant quantum
cloning.
\end{abstract}


\maketitle

\section{Introduction}\label{sec:secIntro}
Recent development in quantum information have given rise to an
increasing number of applications, for instance, quantum
teleportation, quantum dense coding, quantum cryptography, quantum
logic gates, quantum algorithms and etc~\cite{Galindo2002,Gisin and
Thew,Gisin and Ribordy,rev}. Many tasks in quantum information
processing (QIP) have different properties from the classical
counterpart, for example, the quantum cloning. Classically, we can
duplicate (copy) any bits perfectly. In the quantum case, as shown
by Wooters and Zurek~\cite{Wootters1982}, it is impossible to design
a general machine to clone every state on the Bloch's sphere
perfectly. This is called the no-cloning theorem. But such the
no-cloning theorem~\cite{Wootters1982} only forbids the perfect
cloning.  As shown by Bu\v{z}ek and Hillery, approximate cloning of
an unknown quantum state is possible. They proposed a type of
Universal Quantum Copying Machine~\cite{Buzek1996} (UQCM) that
clones all the state on the Bloch's sphere with the same optimal
fidelity~\cite{Gisin1997,Brub1998,Gisin1998}. Subsequently, some
researches have extended the UQCM to $N$ inputs to $M$ outputs and
to d-level system~\cite{Gisin1997,Werner1998,Keyl1999}. Furthermore,
studies have also been done on quantum cloning with prior
information about unknown
state~\cite{Brub2000,phase1,phase2,buc,Fiu2003,Cerf2002,Fan2003,Cerf
and Durt2002,Durt2003}, the example is the cloning of phase
covariant state~\cite{Brub2000} or unknown equatorial
state~\cite{Fan2003} given by
$$\ket{\psi}=\frac{1}{\sqrt{2}}\left(\ket{0}+e^{i\phi}\ket{1}\right).$$
It has already been proven that the above state can be cloned with
the optimal fidelity
$F=\frac{1}{2}[1+\frac{1}{\sqrt{2}}]$~\cite{Fan2001} and the
fidelity is higher than UQCM's. This is to say, if we already have
some prior information about the unknown state, we can design a
better copying machine for the state. The
result~\cite{Brub2000,phase1,phase2,Fan2003} was subsequently
extended to more general case~\cite{Karimipour2002,phase4} and
experimentally demonstrated \cite{Fiu2003,phase4,phase5}.

The UQCM and the phase covariant cloning do not subsume each other,
because one cannot be regarded as a special case of the other. In
real applications of the quantum information system, we sometimes
have access only to pure states distributed on a specific surface on
a Bloch sphere. In this article, we study out such general situation
in which the states are distributed between two latitudes on a Bloch
sphere. Our result unifies the prior results pertaining to UQCM and
phase covariant cloning: in particular, one could bring the two
latitudes to the poles for UQCM or set the two latitudes together
for phase covariant cloning.

To this end, we consider the following state:
\begin{equation}
  \ket{\psi}=\cos{\frac{\theta}{2}}\ket{0}+
  \sin{\frac{\theta}{2}}e^{i\phi}\ket{1}
\end{equation}
where $\phi\in[0,2\pi]$ and $\theta_{1}\leq\theta\leq\theta_2$. The
states we considered here are uniformly distributed between two
latitudes on the Bloch sphere. When $\theta_{1}=0$ and
$\theta_{2}=\pi$, we get the situation of the UQCM. When
$\theta_{1}=\theta_{2}=\frac{\pi}{2}$, it is the phase covariant
cloning. In this way, results of the UQCM and the phase covariant
cloning can be unified: they are recovered as special cases of our
QCM. Contrary to general perception that we can get a better QCM, we
point out that this view may not always be true.

This paper is arranged as follows: After introducing some results
concerning UQCM~\cite{Buzek1996} and phase covariant
cloning~\cite{Brub2000,Fan2003,Fan2001,Karimipour2002}, we formulate
our problem in section \uppercase\expandafter{\romannumeral2} and
present analytical results to the situation. In section
\uppercase\expandafter{\romannumeral3}, we make detailed discussions
about our 1 $\to$ 2 QCM and also a qualitative discussion about the
situation of $1\to N$ and $M\to N$. We end the paper with some
concluding remarks.

For an arbitrary quantum state on the Bloch's sphere, we can use the
following unitary transformation to get the optimal result for the
cloning:
\begin{eqnarray}
  U: &&
  \ket{0}_{a}\ket{0}_{b}\ket{\uparrow}_{x}\rightarrow
  \sqrt{\frac{2}{3}}\ket{0}_{a}\ket{0}_{b}\ket{\uparrow}_{x}+
  \sqrt{\frac{1}{6}}\left(\ket{0}_{a}\ket{1}_{b}+
  \ket{1}_{a}\ket{0}_{b}\right)\ket{\downarrow}_{x}\nonumber \\
  &&
  \ket{1}_{a}\ket{0}_{b}\ket{\uparrow}_{x}\rightarrow
  \sqrt{\frac{2}{3}}\ket{1}_{a}\ket{1}_{b}\ket{\downarrow}_{x}+
  \sqrt{\frac{1}{6}}\left(\ket{0}_{a}\ket{1}_{b}+
  \ket{1}_{a}\ket{0}_{b}\right)\ket{\uparrow}_{x}.
\end{eqnarray}

For the state $\ket{\psi}=\alpha\ket{0}+\beta\ket{1}$, after
operated by the cloning operation U, we can get the density matrices
$\rho_{a}$ and $\rho_{b}$ by taking partial trace. We then define
the cloning fidelity $F=\bra{\psi}\rho_{a}\ket{\psi}$. For the case
of 1 to 2 UQCM, it can be proved that
$F=\frac{5}{6}$~\cite{Buzek1996}.

For the phase covariant cloning there already exists a method of
adjusting a parameter in the UQCM to get a better cloning
fidelity~\cite{Karimipour2002}.

\section{Quantum cloning machine for a qubit between two latitudes on the Bloch
sphere}\label{sec:secQCM}

The state we wish to clone can be written as
\begin{equation}\ket{\psi}=\cos{\frac{\theta}{2}}\ket{0}+
\sin{\frac{\theta}{2}}e^{i\phi}\ket{1}\end{equation} where
$\phi\in[0,2\pi]$ and
\begin{equation}\theta_1\leq\theta\leq\theta_2.\end{equation}
This is to say, the states we considered here are uniformly
distributed in a belt between two latitudes on the Bloch sphere. We
assume the following unitary transformation for our QCM:
\begin{eqnarray}\label{eq:eqGQCM}
  U: &&
  \ket{0}_{a}\ket{0}_{b}\ket{\uparrow}_{x}\rightarrow
  \cos{\alpha}\ket{0}_{a}\ket{0}_{b}\ket{\uparrow}_{x}+
  \sin{\alpha}\ket{\xi^{+}}_{ab}\ket{\downarrow}_{x}\nonumber \\
  &&
  \ket{1}_{a}\ket{0}_{b}\ket{\uparrow}_{x}\rightarrow
  \cos{\beta}\ket{1}_{a}\ket{1}_{b}\ket{\downarrow}_{x}+
  \sin{\beta}\ket{\xi^{+}}_{ab}\ket{\uparrow}_{x}
\end{eqnarray}
where $\ket{\xi^{+}}$ is defined as
$\ket{\xi^{+}}=\frac{1}{\sqrt{2}}\left(\ket{01}_{ab}+
\ket{10}_{ab}\right)$, with $\alpha$ and $\beta$ being parameters
that we want to determined. We restrain ourselves only to a
'symmetric' transformation and we prove its optimality below.

After transformation by the unitary operation U, we can get the
following state:
\begin{eqnarray}
  \ket{\psi_{a}}\ket{0}_{b}\ket{\uparrow}_{x}
  & \rightarrow &
  \cos{\frac{\theta}{2}}\cos{\alpha}\ket{00}_{ab}\ket{\uparrow}_{x}+
  \sin{\alpha}\cos{\frac{\theta}{2}}\ket{\xi^{+}}_{ab}\ket{\downarrow}_{x}
  \nonumber \\
  & &
  +\sin{\frac{\theta}{2}}\cos{\beta}e^{i\phi}\ket{11}_{ab}\ket{\downarrow}_{x}+
  \sin{\frac{\theta}{2}}\sin{\beta}e^{i\phi}\ket{\xi^{+}}_{ab}\ket{\uparrow}_{x}.
\end{eqnarray}
By taking partial trace, we can calculate the reduced density
matrices $\rho_{a}$ and $\rho_{b}$ of particle a and b respectively.
\begin{eqnarray}
  \rho_{a}=\rho_{b}
  &=&
  \left(\frac{1}{\sqrt{2}}\sin{\beta}\sin{\frac{\theta}{2}}\right)^2
  \oprod{0}{0}+\left(\frac{1}{\sqrt{2}}\sin{\alpha}\cos{\frac{\theta}{2}}\right)^2
  \oprod{1}{1} \nonumber \\
  & &
  + \left(\cos{\frac{\theta}{2}}\cos{\alpha}\ket{0}+
  \frac{1}{\sqrt{2}}\sin{\frac{\theta}{2}}\sin{\beta}e^{i\phi}\ket{1}\right)
  \left(\cos{\frac{\theta}{2}}\cos{\alpha}\bra{0}+
  \frac{1}{\sqrt{2}}\sin{\frac{\theta}{2}}\sin{\beta}e^{-i\phi}\bra{1}\right)
  \nonumber \\
  & &
  +\left(\sin\frac{\theta}{2}\cos{\beta}e^{i\phi}\ket{1}+
  \frac{1}{\sqrt{2}}\cos{\frac{\theta}{2}}\sin{\alpha}\ket{0}\right)
  \left(\sin\frac{\theta}{2}\cos{\beta}e^{-i\phi}\bra{1}+
  \frac{1}{\sqrt{2}}\cos{\frac{\theta}{2}}\sin{\alpha}\bra{0}\right).
\end{eqnarray}
With the density matrices of the subsystem, we can get the fidelity
\begin{eqnarray}
  F
  &=&
  \bra{\psi}\rho_{a}\ket{\psi} \nonumber \\
  &=&
  \cos^{4}{\frac{\theta}{2}}\left(\frac{1}{2}+\frac{1}{2}\cos^{2}{\alpha}\right)
  +\sin^{4}{\frac{\theta}{2}}\left(\frac{1}{2}+\frac{1}{2}\cos^{2}{\beta}\right)
  +\frac{1}{8}\sin^{2}{\theta}\left(\sin^{2}{\alpha}+\sin^{2}{\beta}\right)
  +\frac{\sqrt{2}}{4}\sin^{2}{\theta}\sin(\alpha+\beta).
\end{eqnarray}
Averaging the fidelity over all possible angles $\theta$, we
have~\cite{Gisin1997}
\begin{eqnarray}
  \bar{F}
  &=&
  \frac{\int_{\theta_{1}}^{\theta_{2}}{F\sin{\theta}d{\theta}}}
  {\int_{\theta_{1}}^{\theta_{2}}{\sin{\theta}d{\theta}}}
  \nonumber\\
  &=&
  \frac{1}{2}+\frac{1}{6}K-P\sin(\alpha+\beta)-Q\sin^{2}{\alpha}-R\sin^{2}{\beta}
\end{eqnarray}
where
\begin{equation}\label{eq:eqPara}
  \left\{
  \begin{array}{l}
    K=\cos^{2}{\theta_{2}}+\cos{\theta_1}\cos{\theta_2}+\cos^{2}{\theta_2}
    \\
    P=\frac{\sqrt{2}}{12}K-\frac{\sqrt{2}}{4} \\
    Q=\frac{1}{12}K+\frac{1}{8}\left(\cos{\theta_1}+\cos{\theta_2}\right)
    \\
    R=\frac{1}{12}K-\frac{1}{8}\left(\cos{\theta_1}+\cos{\theta_2}\right)
  \end{array}
  \right.
\end{equation}
and $K,P,Q,R$ are constants with given $\theta_1$ and $\theta_2$. In
order to get the maximum of $\bar{F}$, we do a partial
differentiating $\bar{F}$ with respect to $\alpha$ and $\beta$. For
maximum $\bar{F}$, the parameters $\alpha$ and $\beta$ with the
optimal QCM should satisfy the following equations:
\begin{equation}\label{eq:eqQCM}
  \left\{
  \begin{array}{l}
    P\cos(\alpha+\beta)+Q\sin(2\alpha)=0\\
    P\cos(\alpha+\beta)+R\sin(2\beta)=0.
  \end{array}
  \right.
\end{equation}

\subsection{Solution
of $\alpha$ and $\beta$ with maximum $\bar{F}$.}

With the formulation above, we can now seek the  solution of
$\alpha$ and $\beta$ with maximum $\bar{F}$. Consider the situation
in which the state cover the whole Bloch sphere. For this situation,
$\theta_{1}=0$ and $\theta_2=\pi$, and we have $K=1$,
$P=-\frac{\sqrt{2}}{6}$ and $Q=R=\frac{1}{12}$. We can also solve
the equations~\eqref{eq:eqQCM} to get
$\cos{\alpha}=\cos{\beta}=\sqrt{\frac{2}{3}}$. This is the well
known result in UQCM.

Now the general situation, we can easily get
\begin{equation}
  \cos(\alpha+\beta)\left[2QR\sin(\alpha-\beta)+P(R-Q)\right]=0.
\end{equation}

If it satisfies that
\begin{equation}\label{eq:eqcond}
  \left|\frac{P(Q-R)}{2QR}\right|\leq 1
\end{equation}
we have
\begin{equation}
  \sin{(\alpha-\beta)}=\frac{P(Q-R)}{2QR}.
\end{equation}
Then we can get the following solution
\begin{equation}\label{eq:eqsolution}
  2\alpha=\arcsin\left[\frac{P(Q+R)}{S}\right]+
  \arcsin\left[\frac{P(Q-R)}{2QR}\right], \quad
  2\beta=\arcsin\left[\frac{P(Q+R)}{S}\right]-
  \arcsin\left[\frac{P(Q-R)}{2QR}\right]
\end{equation}
where $S=-\sqrt{4QRP^2+4Q^{2}R^{2}}$. If
$\left|\frac{P(Q-R)}{2QR}\right|>1$, we can get
$\cos(\alpha+\beta)=0$ and $\sin{2\alpha}=0, \sin{2\beta}=0$. Since
$P\leq 0$, there are only two cases
\begin{enumerate}
  \item If $|\theta_{1}-\frac{\pi}{2}|\geq
  |\theta_{2}-\frac{\pi}{2}|$, we have
  $\alpha=0,\beta=\frac{\pi}{2}$ and
  $\bar{F}=\frac{1}{2}+\frac{1}{6}K-P-R$;
  \item if $|\theta_{1}-\frac{\pi}{2}|<
  |\theta_{2}-\frac{\pi}{2}|$, we have
  $\alpha=\frac{\pi}{2},\beta=0$ and
  $\bar{F}=\frac{1}{2}+\frac{1}{6}K-P-Q$.
\end{enumerate}

In summary, the mean fidelity of our QCM is:
\begin{equation}\label{eq:eqMF}
  \bar{F}=\left\{
  \begin{array}{ll}
    \frac{1}{2}+\frac{1}{6}K-P\sin(\alpha+\beta)-
    Q\sin^{2}{\alpha}-R\sin^{2}{\beta}, & |T|\leq 1; \\
    \frac{1}{2}+\frac{1}{6}K-P-R, & |T|>1 \, \textrm{and} \,
    |\theta_{1}-\frac{\pi}{2}|\geq |\theta_{2}-\frac{\pi}{2}|; \\
    \frac{1}{2}+\frac{1}{6}K-P-Q, & |T|>1 \, \textrm{and} \,
    |\theta_{1}-\frac{\pi}{2}|< |\theta_{2}-\frac{\pi}{2}|.
  \end{array}
  \right.
\end{equation}
where $T=\frac{P(Q-R)}{2QR}$ and $\alpha, \beta$ are given in
Eq.~\eqref{eq:eqsolution}.

\subsection{Optimization}

Our QCM discussed above has a 'symmetric' form defined by
Eq.~\eqref{eq:eqGQCM}. In this section, we will prove that the
symmetric form is necessary. We consider all possible forms of
quantum cloning. If the transformation U of the bases is not in such
a symmetric form, we know that the reduced density matrices of the
particle a and b are not equal to each other. We then get the
different fidelities for particle a and b
$$F_{a}=\bra{\psi}\rho_{a}\ket{\psi}, \quad\quad
F_{b}=\bra{\psi}\rho_{b}\ket{\psi}.$$ For the purpose of the
cloning, we can only define the fidelity as follows
\begin{equation}
  F_{U}=\min(F_{a},F_{b})
\end{equation}
Then we can construct another quantum cloning machine Q to satisfy:
\begin{equation}
  \rho_{a}^{'}=\rho_{b}, \quad \quad \rho_{b}^{'}=\rho_{a}
\end{equation}
where $\rho_{a}$ and $\rho_{b}$ are the reduced density matrices of
two particles a and b.

If we use U and Q with probability $\frac{1}{2}$ to copy the state,
the reduced density matrices of two particles will be the same
$\frac{1}{2}(\rho_{a}+\rho_{b})$. In this situation, the cloning
fidelity $F_{Q}=\bra{\psi}\frac{1}{2}(\rho_{a}+\rho_{b})\ket{\psi}$.
We have $F_{Q}\geq F_{U}$. Finally we get a symmetric form. It can
also be said that for every possible cloning, we can always find a
'symmetric' cloning transformation that is optimal. So we need only
consider the form given by Eq.~\eqref{eq:eqGQCM}. After getting the
solution to this symmetric situation, we find the optimal QCM.

\section{Some Discussion about our QCM }

Given $\theta_{1}$ and $\theta_{2}$, we can calculate the optimal
fidelity by using Eq.~\eqref{eq:eqMF}. Fig.1 presents all the
situation with states uniformly distributed in any belt on the Bloch
space. With observation from Fig.1 and simple derivation, we arrive
at the following results:
\begin{enumerate}
  \item If $\theta_1=0,\theta_2=\pi$, it is the situation of the UQCM
  and the optimal fidelity is $F=\frac{5}{6}$ which corresponds to
  points B$_{1}$ or B$_{2}$ in Fig.1;
  \item If $\theta_{1}=\theta_{2}=\frac{\pi}{2}$, we encounter the
  situation of Phase-covariant QCM  and the optimal fidelity is
  $F=\frac{1}{2}(1+\frac{1}{\sqrt{2}})$ which corresponds to the
  point C in Fig.1;
  \item Fixed one latitude of the belt, we can set $\theta_1$ to be
  constant without closing any generality, the minimum fidelity
  will be get at the point with $\theta_2=\pi-\theta_1$. For
  example, fixed $\theta_1=\frac{\pi}{4}$, Fig.2 draw the optimal
  fidelity with $\theta_2\in[\frac{\pi}{4},\pi]$. The minimum
  optimal fidelity obtained when
  $\theta_{2}=\pi-\theta_{1}=\frac{3\pi}{4}$. Contrary to one's
  intuition, the fidelity of cloning can rise with the area of where
  t he unknown input state is on.
\end{enumerate}

\begin{figure}[h]
  \begin{center}
  \includegraphics[width=115mm]{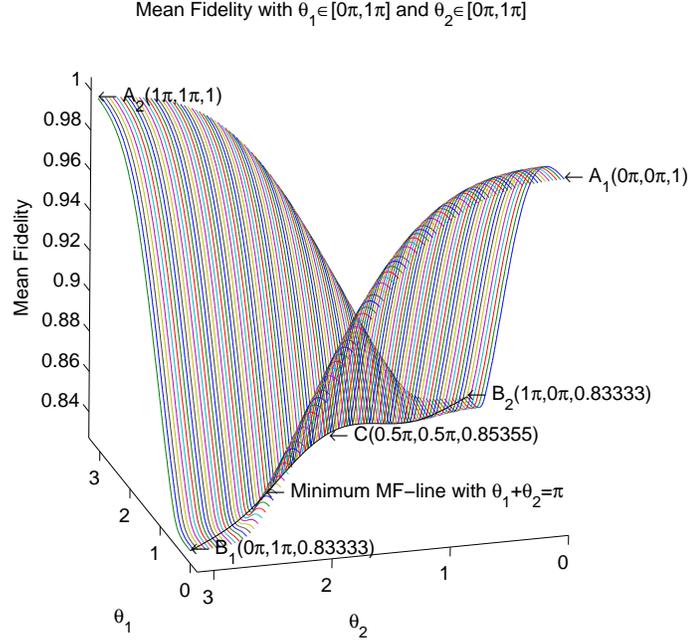}
  \caption{The optimal fidelity of 1 to 2 cloning for states between any
  tow latitudes of Bloch space. Point B$_{1}$ and B$_{2}$ correspond to the
  situation of UQCM. Point C corresponds to the situation of
  Phase-covariant QCM. The bottom line corresponds to the situation
  of $\theta_{1}+\theta_{2}=\pi$.}
  \end{center}
  \label{fig:figTotal}
\end{figure}

\begin{figure}[h]
  \begin{center}
  \includegraphics[width=100mm]{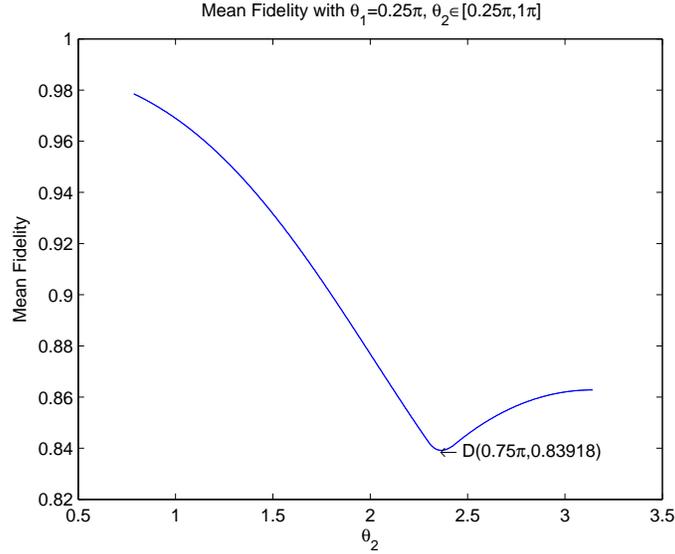}
  \caption{The optimal fidelity with $\theta_{1}=\frac{\pi}{4},
  \theta_{2}\in [\frac{\pi}{4},\pi]$. Point D corresponds to the
  minimum optimal fidelity with $\theta_{2}=\frac{3\pi}{4}$.}
  \end{center}
  \label{fig:figMin}
\end{figure}

In general, we sometimes will encounter the problem of multiple
cloning, i.e. $1\to N$ and $M\to N$. Here we discuss the results for
$1\to 2$ cloning to case of $1\rightarrow N$ and $M\rightarrow N$
qualitatively.

For the states $\ket{\psi}=\cos{\frac{\theta}{2}}\ket{\uparrow}+
\sin{\frac{\theta}{2}}e^{i\phi}\ket{\downarrow}$, where
$\phi\in[0,2\pi]$ and $\theta_1\leq\theta\leq\theta_2$.

We can assume~\cite{Fan2001}
\begin{eqnarray}
  U_{1,N}\ket{\uparrow}\otimes R
  &=&
  \sum_{j=0}^{N-1}{a_{j}\ket{(N-j)\uparrow,j\downarrow}\otimes
  R^{j}} \nonumber \\
  U_{1,N}\ket{\downarrow}\otimes R
  &=&
  \sum_{j=0}^{N-1}{b_{M-1-j}\ket{(N-1-j)\uparrow,(j+1)\downarrow}\otimes
  R^{j}} \nonumber
\end{eqnarray}
where $R$ and $R_j$ are the auxiliary quantum system.

We know that the parameters of particle a and b are not completely
independent. We must let the reduced density matrices of $N$
particles to be the same form in order to achieve optimality, so we
can assume that the cloning transformation of bases take a symmetric
form.

Using the same method and defining the fidelity as
$F=\bra{\psi}\rho_{a}\ket{\psi}$, we get
\begin{equation}
  \bar{F}=\frac{\int_{\theta_1}^{\theta_2}{F\sin{\theta}d{\theta}}}
  {\int_{\theta_1}^{\theta_2}{\sin{\theta}d{\theta}}}.
\end{equation}
and we calculate the partial derivative es of the free parameters
$a_{j}$ and $b_{j}$ and set them to zero, getting the following
equations:
\begin{equation}
  \frac{\partial{\bar{F}}}{\partial{a_j}}=0,\quad
  \frac{\partial{\bar{F}}}{\partial{b_j}}=0. \quad \quad
  (j=0,1,\cdots,N-1)
\end{equation}
These equations are high-order multi-variant equations and in
general the higher-order equations has no analytical solution. A
similar result can be obtained for the $M\rightarrow N$ situation.

\section{concluding remark}
In summary, we have presented the quantum cloning machine for qubits
uniformly distributed on a belt between two latitudes of the Bloch
sphere. Previous results regards $1\rightarrow 2$ cloning of both
the universal cloning and the phase covariant cloning can unified
into a single formulism. So that the previous results are recovered
as special case of the current results.

{\bf Acknowledgement:} This work was supported in part by the
National Basic Research Program of China grant No. 2007CB907900 and
2007CB807901, NSFC grant No. 60725416 and China Hi-Tech program
grant No. 2006AA01Z420.


\begin{thebibliography}{99}
\bibitem{Galindo2002}
A. Galindo and M.A. Mart\'{i}n-Delgado, Rev. Mod. Phys. 74, 347
(2002)
\bibitem{rev}X.-B. Wang, T. Hiroshima, A. Tomita, and M. Hayashi,
Phys. Rep. 448 (2007)
\bibitem{Gisin and Thew}
N. Gisin and R. Thew, quant-ph/0703255
\bibitem{Gisin and Ribordy}
N. Gisin, G. Ribordy, W. Tittel and H. Zbinden, Rev. Mod. Phys. 74,
145(2002)
\bibitem{Wootters1982}
W.K. Wootters and W.H. Zurek, Nature (London) 299, 802 (1982).
\bibitem{Buzek1996}
V. Bu\v{z}ek and M. Hillery, Phys. Rev. A 54, 1844 (1996).
\bibitem{Gisin1997}
N. Gisin and S. Massar, Phys. Rev. Lett. 79, 2153 (4) (1997).
\bibitem{Brub1998}
D. Bru{\ss}, D.P. DiVincenzo, A. Ekert, C.A. Fuchs, C. Macchiavello
and J.A. Smolin, Phys. Rev. A 57, 2368(1998)
\bibitem{Gisin1998}
N. Gisin, Phys. Lett. A 242, 1(1998)
\bibitem{Werner1998}
R.F. Werner, Phys. Rev. A 58,1827(1998)
\bibitem{Keyl1999}
M. Keyl and R.F. Werner, J. Math. Phys. 40, 3283(1999)
\bibitem{Brub2000}
D. Bru{\ss}, M. Cinchetti, G.M. D'Ariano and C. Macchiavello, Phys.
Rev. A 62, 012302 (2000).
\bibitem{phase1} G.M. D'Ariano and P. Lo Presti, Phys. Rev. A 64,
042308 (2001)
\bibitem{phase2}
G.M. D'Ariano and C. Macchiavello, Phys. Rev. A 67, 042306 (2003)
\bibitem{buc}F. Buscemi, G.M. D'Ariano and C. Macchiavello
Phys. Rev. A 71, 042327 (2005)
\bibitem{Fiu2003}  J. Fiurasek, Phys. Rev. A 67, 052314 (2003)
\bibitem{phase3}A. Cernoch, L. Bartuskova, J. Soubusta, M. Jezek, J. Fiurasek, and M.
Dusek, Phys. Rev. A 74, 042327 (2006)
\bibitem{phase4}W.-H. Zhang, L.-B. Yu, and L. Ye, Phys. Lett. A 353,
130 (2006) \bibitem{phase5} H.-W. Chen, X.-Y. Zhou, D. Suter, and
J.-F. Du JF, Phys. Rev. A 75, 012317 (2007);  X.-B. Zou and W.
Mathis, Phys. Rev. A 72, 022306 (2005)
\bibitem{Cerf2002}
N.J. Cerf, M. Bourennane, A. Karlsson and N. Gisin, Phys. Rev. Lett.
88,127902(2002)
\bibitem{Fan2003}
H Fan, H Imai, K. Matsumoto and X.B. Wang, Phys. Rev. A 67, 022317
(2003).
\bibitem{Fan2001}
H Fan, K Matsumoto, X.B. Wang and M. Wadati, Phys. Rev. A 65, 012304
(2001).
\bibitem{Cerf and Durt2002}
N.J. Cerf, T. Durt and N. Gisin, J. Mod. Opt. 49,1355(2002)
\bibitem{Durt2003}
T. Durt and B. Nagler, Phys. Rev. A 68, 042323(2003)
\bibitem{Karimipour2002}
V. Karimipour and A.T. Rezakhani, Phys. Rev. A 66, 052111 (2002).

\end{thebibliography}

\end{document}